%% file: paper5.tex
\documentclass[a4paper]{jpconf}
\usepackage{graphicx}
\usepackage{amsmath}
\usepackage{times}
\usepackage{multirow}
\usepackage{color}
\usepackage{array}
\usepackage{citesort}

\def\be{\begin{equation}}
\def\ee{\end{equation}}
\def\degree{$^\circ$}
\def\etal{\it et al.}

\newcolumntype{C}[1]{>{\centering\let\newline\\\arraybackslash\hspace{0pt}}m{#1}}
\newcolumntype{L}[1]{>{\raggedright\let\newline\\\arraybackslash\hspace{0pt}}m{#1}}
\newcommand*{\mytab}{\hspace*{3mm}}%

\begin{document}
\title{{\it Ab initio} density-functional studies of 13-atom Cu and Ag clusters}
\author{Dil K. Limbu$^1$, Michael U. Madueke$^2$, Raymond Atta-Fynn$^2$,
David A. Drabold$^3$ and Parthapratim Biswas$^1$}
\address{$^1$ Department of Physics and Astronomy, The University of 
Southern Mississippi, Hattiesburg, Mississippi 39406, USA}
\address{$^2$ Department of Physics, University of Texas, Arlington, Texas 76019, USA}
\address{$^3$ Department of Physics and Astronomy, Ohio University, Athens, Ohio 45701, USA}
\ead{partha.biswas@usm.edu}

\begin{abstract}
The putative ground-state structures of 13-atom Cu and Ag clusters have been 
studied using {\it ab initio} molecular-dynamics (AIMD) simulations based on the 
density-functional theory (DFT). An ensemble of low-energy configurations, 
collected along the AIMD trajectory and optimized to nearest local 
minimum-energy configurations, were studied.  An analysis of the results 
indicates the existence of low-symmetric bilayer structures as strong 
candidates for the putative ground-state structure of Cu$_{13}$ and Ag$_{13}$ 
clusters.  These bilayer structures are markedly different from a buckled 
bi-planar (BBP) configuration and energetically favorable, by about 0.4--0.5 eV, 
than the latter proposed earlier by others. Our study reveals that the 
structure of the resulting putative global-minimum configuration is 
essentially independent of the nature of basis functions (i.e., plane waves 
vs.\,pseudoatomic orbitals) employed in the calculations, for a given 
exchange-correlation functional. The structural configurations obtained 
from plane-wave-based DFT calculations show a slightly tighter or 
dense first-shell of Cu and Ag atoms than those from local-basis 
functions.  A comparison of our results with 
recent full-potential DFT simulations is presented.
\end{abstract}

\section{Introduction}
Transition-metal (TM) clusters have been studied extensively from 
computational and experimental points of 
view~\cite{Baletto2005,Doye1998,Elliott2009a,Parks1995,Oyanagi2012} due to 
their potential applications in catalysis~\cite{Henry1998}, 
photonics~\cite{Barnes2003} and carbon nanotubes~\cite{Li2004}. 
Owing to the difficulties associated with the experimental 
determination of the structures of small 
clusters~\cite{Baletto2005}, Monte Carlo and 
molecular-dynamics simulations play important roles 
in the structural characterization of small clusters.  While 
computational studies based on classical interatomic potentials 
indicate that the icosahedral structure is the preferred 
ground state of 13-atom Cu and Ag clusters~\cite{Doye1998,Limbu2017a}, 
{\it ab initio} studies based on the density-functional theory (DFT)~\cite{Kohn1965} 
indicate the presence of a few competing structures as 
possible ground-state structures~\cite{Oviedo2002,Chang2004,Longo2006,Sun2008,Piotrowski2010,Chou2013,McKee2017,Chaves2017,Limbu2017}, depending upon the 
types of the basis functions and exchange-correlation (XC) 
functionals employed in the DFT calculations. Furthermore, given 
the computational complexity of {\it ab initio} calculations, 
the results can depend considerably on the total simulation 
time and the method used to sample candidate structures from 
the potential-energy surface (PES) during simulations.  Of 
particular interest among {\it ab initio} studies on 13-atom 
Ag and Cu clusters are the work by Oviedo and 
Palmer~\cite{Oviedo2002} and that by Chang and Chou~\cite{Chang2004}, 
using the first-principles density-functional code 
{\sc Vasp}~\cite{Vasp1996}. The former indicated the presence 
of a few `amorphous-like' low-energy isomers with bilayer 
structures, whereas the latter concluded the existence of 
a buckled bi-planar (BBP) structure as a putative ground 
state of 13-atom Ag and Cu clusters.  A similar conclusion 
was reached by Longo and Gallego~\cite{Longo2006}, who 
employed pseudoatomic orbitals as basis functions in 
their calculations, using the local-basis DFT code {\sc Siesta}~\cite{Siesta2002}. 
Recently, Chaves {\etal} have reported bilayer structures of 
13-atom Cu and Ag clusters using full-potential density-functional 
calculations.  However, a number 
of Gaussian-orbital-based and plane-wave-based DFT 
studies~\cite{Yang2006,Chaves2014,Wang2007,Chen2013} 
reported different structures for Cu$_{13}$ and Ag$_{13}$ clusters. 
Toward this end, the aim of the current study is to present 
results from {\it ab initio} molecular-dynamics simulations, 
lasting for a few hundreds of picoseconds and sampling 
structural configurations from {\it ab initio} 
potential-energy surfaces (PES), using both plane-wave and 
local-basis DFT calculations.  

\section{Computational Method}
Spin-polarized DFT~\cite{Kohn1965} calculations 
were performed within the Perdew-Burke-Ernzerhof (PBE) formulation~\cite{PBE} 
of the generalized gradient approximation (GGA) using the DFT code {\sc Siesta}~\cite{Siesta2002}. 
Pseudoatomic-orbital double-zeta basis with polarization (DZP) functions and norm-conserving 
Troullier-Martins pseudopotentials~\cite{TM1991} were employed. All calculations were 
performed using a cubic simulation cell of size 30 {\AA}. The large simulation 
cell permitted us to sample the Brillouin zone using the {\bf $\Gamma$}-point 
only.  A large number of low-energy configurations of Ag$_{13}$ 
and Cu$_{13}$ clusters were generated from {\it ab initio} molecular-dynamics simulations 
in the canonical ensemble followed by geometry optimizations. 
The simulation procedure is as follows. A randomly generated 13-atom 
cluster placed in the box.  The system was initially 
equilibrated at 1500 K for 20 ps. The temperature was then 
reduced gradually from 1500 K to 300 K for 240 ps. A total 
of 40 low-energy structures were collected in the temperature range 
of 300 K to 500 K during the AIMD simulations; the geometries 
of the collected structures were then optimized by minimizing 
their respective total energies. The optimizations were based 
on the conjugate-gradient (CG) algorithm; a structure is considered 
to be fully optimized if the force on each atom is less than 
0.005 eV/{\AA}.  The stability of each structure was further 
examined by minimizing the total energy using the more accurate 
plane-wave-based DFT method as implemented in the DFT 
code {\sc Vasp}~\cite{Vasp1996}.

\section{Results and Discussions}
\begin{figure}[ht]
\begin{center}
\includegraphics[angle=-90, width=0.4\textwidth]{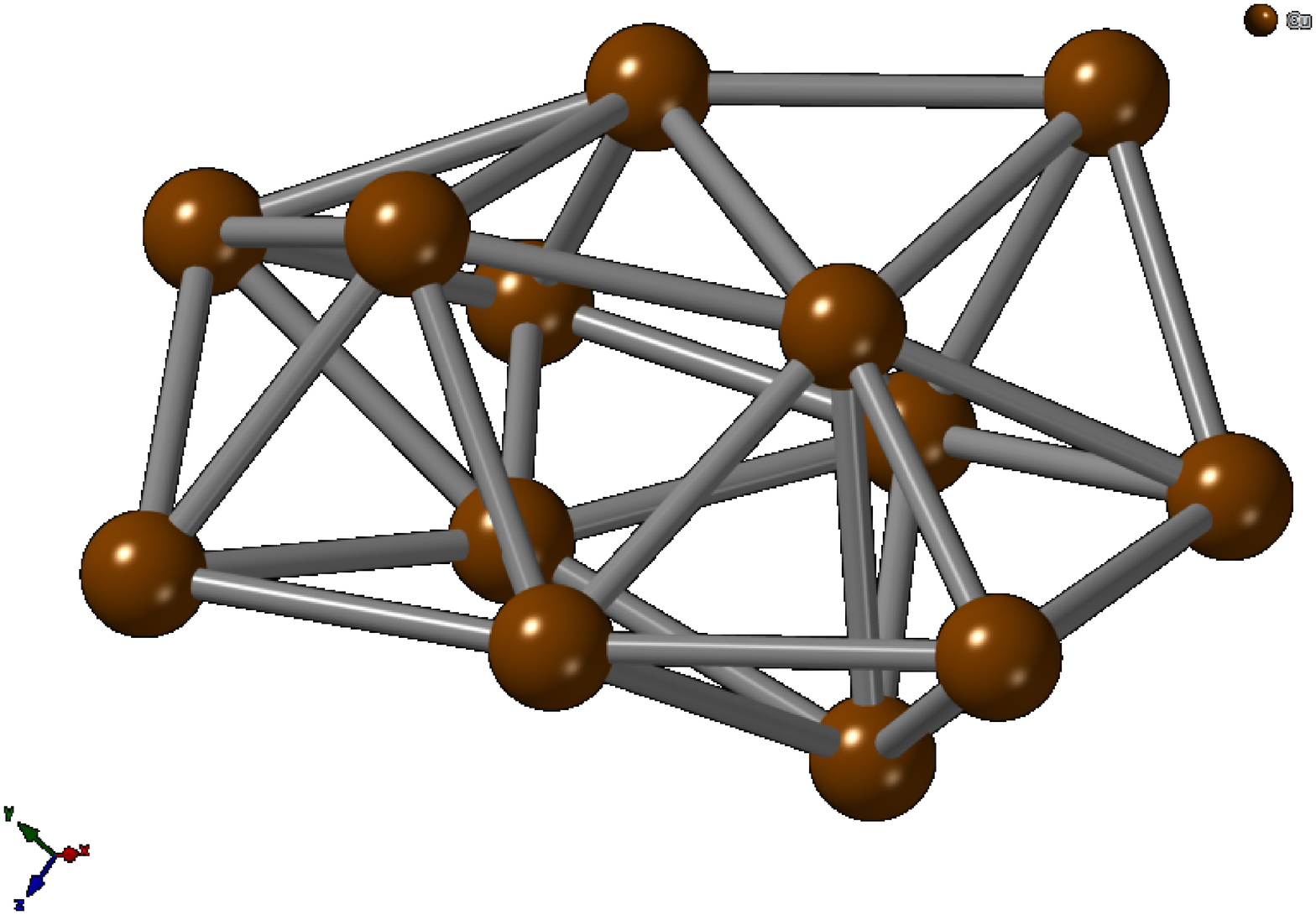}
\includegraphics[angle=-90, width=0.4\textwidth]{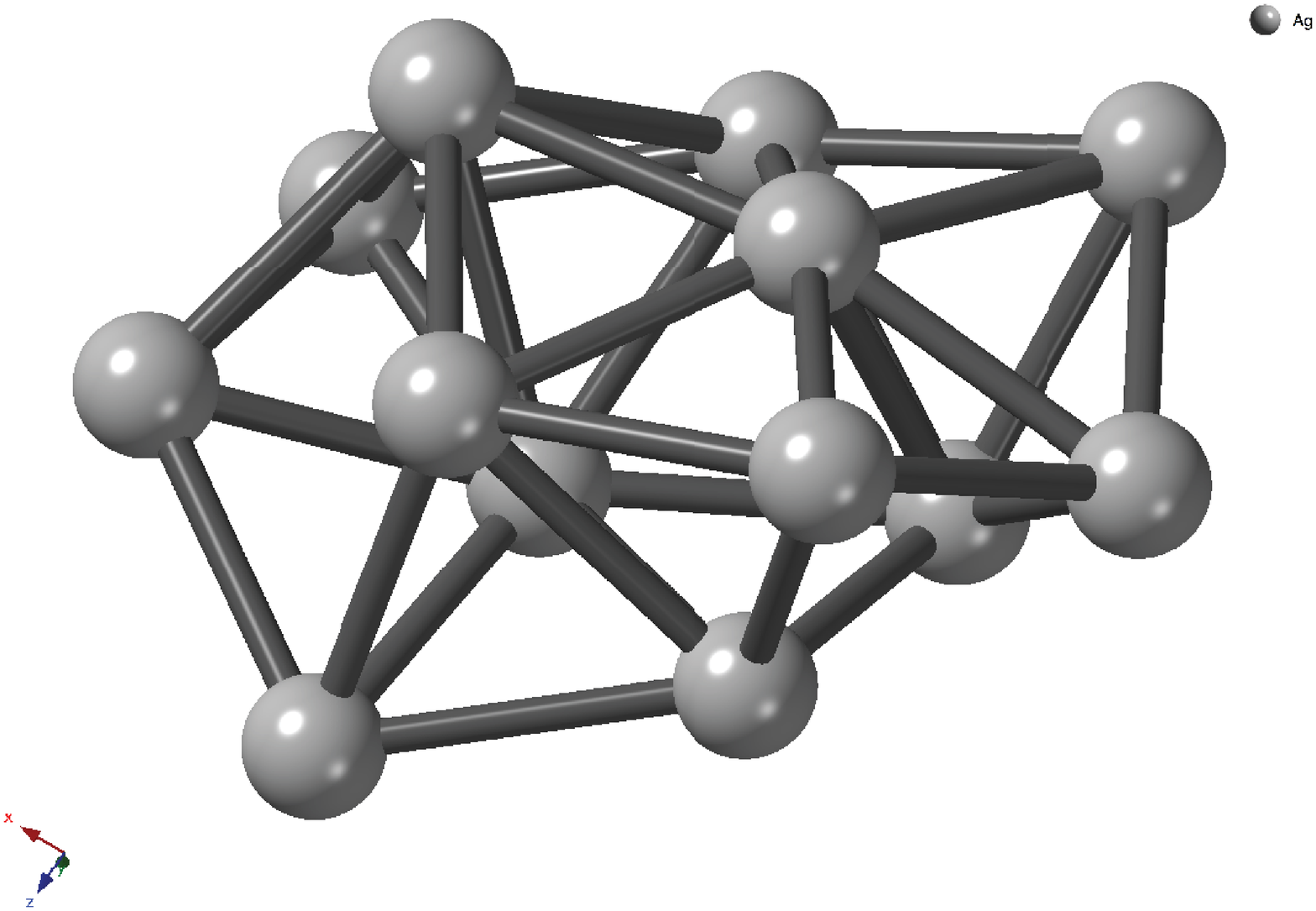}
\caption{\label{3D}
{\small 
The structure of putative global minimum of Cu$_{13}$ (left) 
and Ag$_{13}$ (right) obtained from {\it ab initio} simulations, 
using {\sc Siesta} and {\sc Vasp}. The maximum nearest-neighbor 
distance in each case corresponds to the value obtained from 
the respective pair-correlation function.   }}
\end{center}
\end{figure}

As mentioned earlier, an ensemble of low-energy structural configurations, 
collected during the course of {\it ab initio} molecular-dynamics 
simulations, constitutes a set of candidate structures 
for determining the putative ground-state 
configuration of 13-atom Ag and Cu clusters. Further relaxation of these 
structures, using the plane-wave density-functional code {\sc Vasp}, 
provides the final structure for Ag$_{13}$ and Cu$_{13}$ clusters.  
The potential energy of the putative global minimum for 
Ag and Cu clusters is listed in Table \ref{TAB1}, with respect 
to the potential energy of the corresponding icosahedral structure 
obtained under identical conditions. 
The bilayer nature of these structures is evident 
from Fig.\,\ref{3D}, where a three-dimensional ball-and-stick model 
of the structures are presented.  The results are consistent with the recent 
study by Chaves {\it et al.}\,\cite{Chaves2017}, where similar bilayer 
structures of 13-atom Cu and Ag clusters were reported using DFT calculations. 
\begin{table}[ht]
\caption{\label{TAB1} Total-energy differences (in eV) for 13-atom 
Cu and Ag clusters from their icosahedral counterpart, using 
{\sc Siesta} and {\sc Vasp}.  
}
\begin{center}
\begin{tabular}{L{20mm} C{16mm} C{16mm} C{16mm} C{16mm}}
\br
\mytab $\multirow{2}{*}{Symmetry}$ &\multicolumn{2}{c}{Cu$_{13}:\Delta$E (eV)} &\multicolumn{2}{c}{Ag$_{13}:\Delta$E (eV)}  \\
\cline{2-5}
  &{\small{\sc Siesta}} & {\small{\sc Vasp}} & {\small{\sc Siesta}} & {\small{\sc Vasp}} \\
 \br
\mytab ICO & 0.0 & 0.0 & 0.0 & 0.0 \\
\mytab BBP & -0.460 & -0.462 & -0.514 & -0.785  \\
\mytab Bilayer & -0.926 & -0.972 & -0.934 & -1.231 \\
\mytab Bilayer$^\star$ &\multicolumn{2}{c}{-1.014} &\multicolumn{2}{c}{-1.300}\\
\br
\multicolumn{5}{l}{$^\star$\footnotesize{From Ref.\,\cite{Chaves2017}}}
\end{tabular}
\end{center}
\end{table}

\begin{table}[ht]
\caption{\label{TAB2} Average bond length ($d_{av}$) and 
coordination number ($C_{av}$) for Cu and Ag clusters}
\begin{center}
\begin{tabular}{lcccc}
\br
System &\multicolumn{2}{c}{Cu$_{13}$} &\multicolumn{2}{c}{Ag$_{13}$}  \\
\br
\multirow{2}{*}{Symmetry} & $d_{av}$ ({\AA}) & $C_{av}$ & $d_{av}$ {(\AA}) & $C_{av}$ \\
  &{\small{\sc Siesta}({\sc Vasp})} &{\small{\sc Siesta}({\sc Vasp})} &{\small{\sc Siesta}({\sc Vasp})} &{\small{\sc Siesta}({\sc Vasp})} \\
 \br
ICO & 2.612 (2.500) & 6.395 (6.396) & 3.037 (2.894) & 6.396 (6.396)  \\
BBP & 2.563 (2.454) & 5.469 (5.463) & 2.982 (2.838) & 5.468 (5.460)  \\
Bilayer & 2.570 (2.460) & 5.702 (5.701) & 2.994 (2.845) & 5.709 (5.647) \\
Bilayer$^\S$ & 2.459 & 5.699 & 2.835 & 5.654 \\
\br
\multicolumn{5}{l}{$^\S$\footnotesize{From Ref.\,\cite{Chaves2017}}}
\end{tabular}
\end{center}
\end{table}

\begin{figure}[ht]
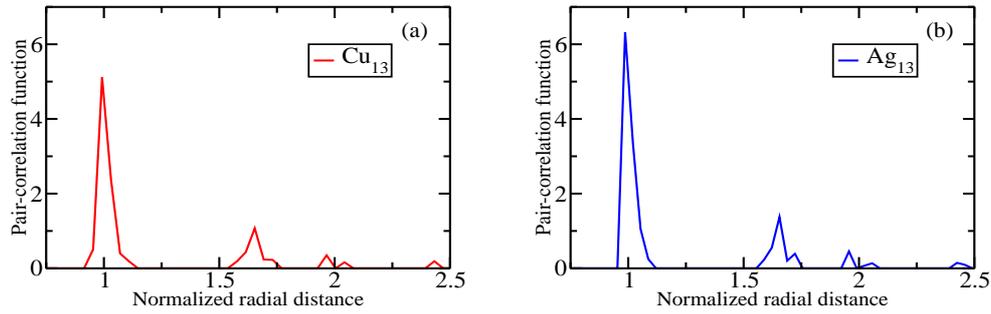

\begin{center}
\includegraphics[width=60mm,height=40mm]{rdfcu.eps}\hspace{8mm}
\includegraphics[width=60mm,height=40mm]{rdfag.eps}
\caption{\label{rdf} 
{\small 
The pair-correlation function of a) Cu$_{13}$ and b) Ag$_{13}$ 
clusters.  For comparison, the radial distances are normalized 
by the corresponding average bond length, $d_{av}$, from {\sc Vasp}, 
as reported in Table \ref{TAB2}.
}}
\end{center}
\end{figure}

\begin{figure}[ht]
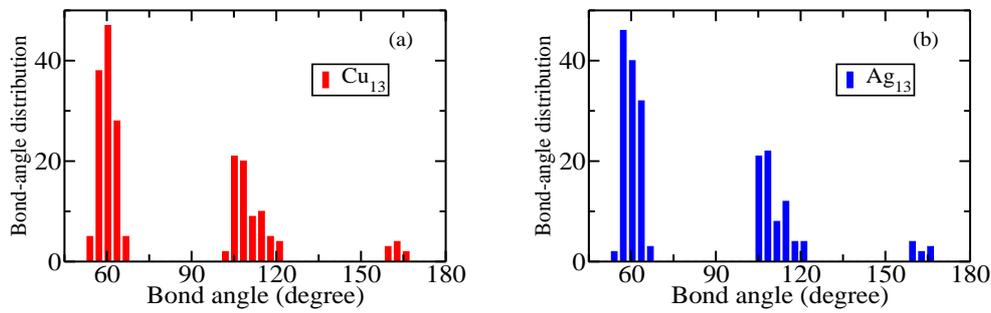

\begin{center}
\includegraphics[width=60mm,height=40mm]{badcu.eps}\hspace{8mm}
\includegraphics[width=60mm,height=40mm]{badag.eps}
\caption{\label{bad}
{\small 
The bond-angle distributions for a) Cu$_{13}$ and b) Ag$_{13}$ clusters. 
}}
\end{center}
\end{figure}

\begin{figure}[ht]
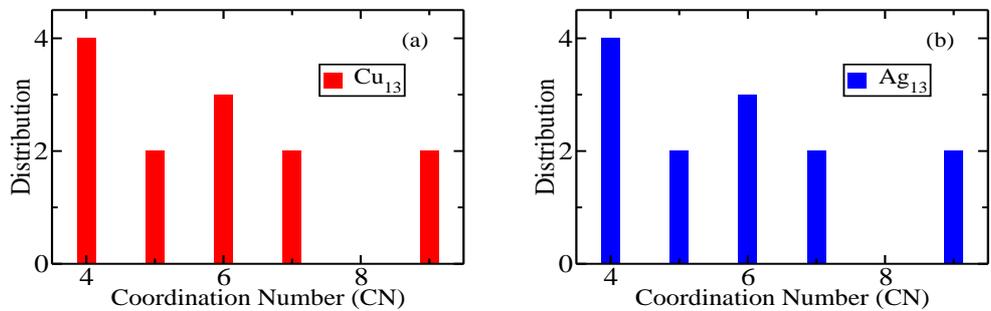

\begin{center}
\includegraphics[width=60mm,height=40mm]{cncu.eps}\hspace{8mm}
\includegraphics[width=60mm,height=40mm]{cnag.eps}
\caption{\label{cn}
{\small 
The distribution of atomic-coordination numbers in a) Cu$_{13}$ and b) Ag$_{13}$ 
clusters. The distribution of the atoms in the first shell is identical 
as far as the bond angles are concerned.
}}
\end{center}
\end{figure}

In order to characterize the three-dimensional distribution of the 
atoms in Ag$_{13}$ and Cu$_{13}$ clusters, we have computed the 
radial and bond-angle distributions, and the average bond length 
(${\bar d_i}$) and the effective coordination number ($C_i$) 
of an atom at site $i$.  The average bond length of an atom 
at site $i$ is given by, 
\be
{\bar d_i} = \sum_j d_{ij}\, p_{ij}, \quad p_{ij} = \frac{e^{f(d_{ij})}}{\sum_j e^{f(d_{ij})}}, \quad f(d_{ij}) = \left[1-\left(\frac{d_{ij}}{\bar{d_i}}\right)^6\right], 
\label{ebl}
\ee
\noindent 
where $d_{ij}$ is the radial distance between two atoms at sites $i$ and 
$j$. The site-average bond length ($d_{av}$) and the average 
coordination number ($C_{av}$) are given by, 
\be 
d_{av} =\frac{1}{N}\sum_{i=1}^N {\bar d_i}, \quad C_{av} = \frac{1}{N}\sum_{i=1}^N C_i, 
\quad C_i = \sum_{j=1}^N e^{f(d_{ij})}
\label{ecn}
\ee 
Equations (\ref{ebl}) and (\ref{ecn}) permit us to calculate the 
average  coordination number and bond length of a cluster 
self-consistently without introducing any arbitrary cutoff 
distance~\cite{Hoppe1979,Chou2013,Chaves2017}. Starting with an 
approximate value of ${\bar d_i}$, one can use Eq.\,(\ref{ebl}) 
to improve the estimate of ${\bar d_i}$ iteratively. 
Table \ref{TAB2} lists the average bond lengths and 
coordination numbers for icosahedral, BBP and 
bilayer structures of 13-atoms Cu and Ag clusters obtained 
from {\sc Siesta} and {\sc Vasp}. The corresponding values computed  
by Chaves {\etal}~\cite{Chaves2017} are also listed for a 
comparison.  An examination of the results from Table \ref{TAB2} suggests 
that the total-energy values obtained from the pseudoatomic-orbital-based 
DFT code {\sc Siesta} slightly overestimates the value of 
total-energy than those obtained from the plane-wave-based 
{\sc Vasp} calculations in our work. This is also reflected in the average 
bond-length of Ag and Cu atoms. The bond lengths obtained from 
{\sc Vasp} relaxations are approximately 4\% shorter than the 
ones calculated from using {\sc Siesta}.
Figure \ref{rdf} shows the pair-correlation functions (PCF) of 
Ag and Cu clusters.  For comparison, the radial 
distances are normalized by the corresponding site-average bond 
length of Ag and Cu atoms, as listed in Table \ref{TAB2}. The 
PCFs look essentially identical except for the height of the 
first peak, which is a reflection of the different average 
coordination number of Ag and Cu clusters. A further characterization 
of a three-dimensional distribution of 
atoms in the clusters is possible by examining the bond-angle 
distribution, as shown in Fig.\,\ref{bad}.  Both the distributions 
exhibit a well-defined peaks near 60{\degree} and 110{\degree}. 
Likewise, the distributions of the effective coordination number 
of atomic sites, $C_i$, of the clusters are shown in Fig.\,\ref{cn}. 

\section{Conclusion}
In this paper, we have used {\it ab initio} molecular-dynamics simulations, 
coupled with geometry optimization,  based on DFT-GGA to predict the 
putative ground state of 13-atom Cu and Ag clusters. An extensive search of 
competitive candidate structures from the {\it ab initio} potential-energy 
surface indicates that a low-symmetry bilayer structure is the most-likely 
candidate for the ground-state structure of Cu$_{13}$ and Ag$_{13}$ clusters. 
A comparison with a number of bilayer structures, optimized with 
pseudoatomic-orbital basis ({\sc Siesta}) and plane-wave basis 
({\sc Vasp}), indicates that the resulting putative global-minimum 
configuration is essentially independent of the nature of basis 
function. The minimum-energy configurations obtained from {\sc Vasp} 
show a slightly shorter average bond length compared to the same 
obtained from {\sc Siesta}.

\ack
This work is partially supported by the U. S. National Science
Foundation under Grants No. DMR 1507166, No. DMR
1507118, and No. DMR 1506836. We acknowledge the use
of computing resources at the Texas Advanced Computing
Center.  

\section*{References}
\input{paper5.bbl}

\end{document}

%% file: paper5.bbl
\providecommand{\newblock}{}